\documentclass[%
 reprint,
 amsmath,amssymb,
 aps,
]{revtex4-2}

\usepackage{graphicx}
\usepackage{dcolumn}
\usepackage{bm}
\usepackage{hyperref}
\usepackage[mathlines]{lineno}

\begin{document}

\preprint{APS/123-QED}

\title{Einstein locality: An ignored core element of quantum mechanics}

\author{Sheng Feng}
 \email{fengsf2a@hust.edu.cn}
\affiliation{%
 School of Electrical and Electronic Information Engineering,\\
 Hubei Polytechnic University, Huangshi, Hubei 435003, China
}%
\affiliation{%
 Center for Fundamental Physics,\\
 Hubei Polytechnic University, Huangshi, Hubei 435003, China
}%




\date{\today}

\begin{abstract}
Quantum mechanics is commonly accepted as a complete theory thanks to experimental tests of non-locality based on Bell's theorem. However, we discover that the completeness of the quantum theory practically suffered from detrimental ignorance of a core element --- Einstein locality. Without this element, important experimental results of relevance could hardly receive full understanding or were even completely misinterpreted. Here we present the discovery with a theory of Einstein locality developed to recover the completeness of quantum mechanics. The developed theory provides a unified framework to account for the results of, e.g.,  Bell experiments (on Bell non-locality) and double-slit experiments with entangled photons (on wave-particle duality). The theory reveals the dynamics of Bell non-locality and the principle of biased sampling in measurement in the double-slit experiments, which otherwise will be impossible tasks without introducing Einstein locality. Worse still, ignorance of this element has caused misinterpretation of observations in the double-slit experiments, leading to perplexing statements of duality violation. Einstein locality also manifests indispensability in theory by its connection to the foundations of other fundamental concepts and topics (e.g., entanglement, decoherence, and quantum measurement) and may advance quantum technology by offering a promising approach to optimizing quantum computing hardware. 
\end{abstract}

\maketitle

{\bf Significance}: Completeness of quantum mechanics has been intensively investigated for over eight decades and widely recognized from the results of Bell experiments (highlighted by 2022 Nobel Prize in Physics). However, this completeness is ruined by practical unawareness of a core element, Einstein locality. Here we develop a theory of Einstein locality to recover the completeness of quantum mechanics. The developed theory exhibits its power by comprehensively accounting for experimental results that otherwise look like just untidy pieces of information of no direct relevance. Moreover, the theory of Einstein locality reveals unknown dynamics or physical principle behind two types of important experiments involving entangled particles. Long-standing confusion about non-locality and claim of wave-particle duality violation are also clarified.


In a renowned paper published in 1935 by Einstein, Podolsky, and Rosen (EPR), the authors raised the completeness issue of quantum mechanics and stated that in a complete theory there is an element corresponding to each element of reality \cite{epr}. Because of practical difficulty in recognizing all elements of reality from the quantum world, to directly prove or disprove the completeness of quantum mechanics is a daunting task that has never been really accomplished. However, for a certified element of a theory, ignorance of this element in practical research will definitely destroy the completeness of this theory, which sets the tone of this study. 

The completeness of quantum mechanics was a main focus in a debate \cite{epr,Bohr1935}, between Bohr and Einstein, that has imposed a far-reaching impact on the development of modern physics. In their paper \cite{epr}, EPR attempted to demonstrate that for two particles in an entangled state, measurement-induced state reductions for one particle should cause an instantaneous state collapse for the other (Einstein non-locality), in accordance with quantum mechanics. By assuming no physical influence faster than the speed of light, they further argued that quantum mechanics is therefore incomplete unless the uncertainty principle is violated. Those involved in the debate struggled to identify a suitable experimental validation for this theory, but a turning point occurred in 1964 when Bell announced his famous inequality \cite{Bell1964}. This provided an excellent quantitative measure of Bell non-locality (i.e., a dependence of the outcome of one event on others occurring at a distance \cite{Bell1964,Brunner2014}), which supports ``physical influences faster than the speed of light" \cite{salart2008,Giustina2013}.

After decades of work leading to successfully conducted Bell experiments \cite{Freedman1972,Aspect1982,salart2008,Giustina2015,Shalm2015,Hensen2015,bigbell2018} and non-locality tests (based on the Bell inequality \cite{Bell1964} or its variants \cite{Clauser1969}), substantial advances have been achieved in the fundamental verification of quantum mechanics. These efforts have furthered our understanding of Bell non-locality \cite{Hirsch2013,Brunner2014,Popescu2014} and have motivated the development of quantum technology used for quantum information processing \cite{nielsen2000,scarani2009,ladd2010,benenti2019,Arute2019,Wu2021,ma2023,Wang2023,Zapatero2023}. However, Bell experiments have left behind a piece of cloud in the sky of quantum physics, i.e., the mysterious nature of Bell non-locality, that still promotes interest among the scientific community \cite{Stapp2009,reid2019,uola2020,Griffiths2020,Bedard2021,Cabello2021}, as a sufficient description has yet to be developed. After all, acquiring a precise definition for "influence" and identifying the dynamic mechanisms of "spooky action at a distance" remain challenging. 

Coincidentally, a second piece of cloud came from double-slit experiments conducted with entangled photons on wave-particle duality \cite{menzel2012,menzel2013,Bolduc2014}. It was claimed \cite{menzel2012} that simultaneous observation of two physical entities, ``wave" and ``particle", is possible by use of correlated photon pairs and this surprising aspect of complementarity was attributed to a special choice of TEM$_{01}$ pump mode. In contrast, other works \cite{Bolduc2014,Leach2016} argued that when biased sampling is allowed in measurement an apparent violation of duality is possible in no need of a TEM$_{01}$ pump mode. But the underlying physical principle behind the biased sampling in measurement is unclear.

In the above experiments, Bell non-locality and wave-particle duality have been studied in parallel lines of inquiry without much intersection. As a result, the identified questions of theoretical importance look just like legacies of little relevance from different types of experiments, which inevitably disrupts enthusiastic efforts from the literature to actively tackle these deep issues. To overcome the predicament one needs to first pin down the fundamental source of problem, which as we discover is ignorance of Einstein locality (i.e., there exists no instantaneous state collapse for a quantum object due to that of its entangled partner \cite{epr,Bell1964}), a core element of quantum mechanics, in interpretation of the corresponding experimental results. The practical unawareness of this element demolished the completeness of quantum mechanics. 

Here we develop a theory of Einstein locality to recover its completeness of quantum mechanics and thereby establish a unified framework to comprehensively account for the results of both the Bell experiments \cite{Freedman1972,Aspect1982,salart2008,Giustina2015,Shalm2015,Hensen2015,bigbell2018} and the double-slit experiments \cite{menzel2012,menzel2013,Bolduc2014}. Importantly, the developed theory can reveal the dynamics of Bell non-locality in the Bell experiments and expose the physical principle governing how sampling was biased in photon measurement in the double-slit experiments. In addition, successful application of the theory based on Einstein locality will bring into light long-standing confusions about non-locality \cite{Giustina2013,reid2019,uola2020} and duality violation \cite{menzel2012,menzel2013,Bolduc2014}; The confusions are warning signs for demolition of the completeness of quantum mechanics caused by practical ignorance of its core element, Einstein locality.

\section*{Theory of Einstein locality}
To retrieve the completeness of quantum mechanics, a theory of Einstein locality is developed as presented below, guided by the following insight into the involved problem: Realistic ignorance of Einstein locality in previous research activities was due to the introduction of Einstein non-locality by EPR in their paper \cite{epr}. As will be proved here, the concept of Einstein non-locality actually came from an extremely subtle mistake in EPR's application of the superposition principle to a wave function for a two-particle system, which constituted a Schmidt decomposition of the eigenfunctions for two operators describing a joint system measurement \cite{epr}.

To illustrate, the superposition principle is considered here for two electrons, $a$ and $b$, on which a joint spin measurement is performed. When in an entangled (singlet) state, the electron spins are perfectly anti-correlated and the total spin angular momentum of the whole system is zero. Suppose the spin value $\hat{\sigma}_z$ is measured for electron $a$ (namely, $\hat{\sigma}_z^a$), where $\hat{\sigma}_z$ is the $z$-component of the Pauli operator. Similarly, the spin value $\hat{\sigma}_n$ is measured for electron $b$ (namely, $\hat{\sigma}_n^b$), in which $\hat{\sigma}_n$ denotes a component of the Pauli operator along an arbitrary $\mathbf{n}$-axis with a polar angle $\alpha$ and an azimuthal angle $\beta$. The wave function $|\phi_c\rangle$ for these two electrons in a singlet state is then given by:
\begin{eqnarray}
|\phi_c\rangle&=& \sum_{p,q}\gamma_{pq}\ |p_{z}\rangle_a  |q_{n}\rangle_b \ ,
\label{eq:singletzn}
\end{eqnarray}
where $p=\pm$, $q=\pm$, $\gamma_{++}=\gamma_{--}^*=2^{-1/2}\sin(\alpha/2)e^{-i\beta}$, $\gamma_{+-}=-\gamma_{-+}=2^{-1/2}\cos(\alpha/2)$, and $|\pm_{z,n}\rangle$ are respectively the eigenstates of the $\hat{\sigma}_{z,n}$ operators (i.e., $\hat{\sigma}_{z}|p_{z}\rangle=p|p_{z}\rangle$ and $\hat{\sigma}_{n}|q_{n}\rangle=q|q_{n}\rangle$). The $|\cdot\rangle_{a,b}$ term represents the states of the labeled particles. Note that equation (\ref{eq:singletzn}) cannot be expressed as a Schmidt decomposition of $|\phi_c\rangle$ for the joint $\hat{\sigma}_z^a\hat{\sigma}_n^b$ measurement unless $\alpha= 0$, which is a crucial yet subtle detail unaddressed by EPR in their paper (see Methods Mark$_1$ for detailed calculations and explanations).

Each term on the right-hand side of equation (\ref{eq:singletzn}) represents one possible result of a joint $\hat{\sigma}_z^a\hat{\sigma}_n^b$ measurement performed on the system, with a probability of $|\gamma_{pq}|^2$, in which a corresponding physical event will not occur before the $\hat{\sigma}_z^a\hat{\sigma}_n^b$ measurement is completed. For instance, there are two possible events in the case that electron $a$ is projected onto the $|+_{z}\rangle_a$ state by a $\hat{\sigma}_z^a$ measurement. The compound system may be 1) projected onto the $|+_{z}\rangle_a  |+_{n}\rangle_b$ state or 2) projected onto the $|+_{z}\rangle_a  |-_{n}\rangle_b$ state. However, none of these potential events will occur until the joint measurement is finished, which implies that electron $b$ will not collapse into the $|+_{n}\rangle_b$ or the $|-_{n}\rangle_b$ states unless a $\hat{\sigma}_{n}^b$ measurement is performed. In other words, electron $b$ will remain in its initial state until it is projected onto a $|q_{n}\rangle_b$ state by a $\hat{\sigma}_{n}^b$ measurement. As such, this particle will not undergo an instantaneous state collapse due to the measurement-induced state collapse of electron $a$, which represents Einstein locality of quantum mechanics (see Methods Mark$_2$ for detailed discussion). 

The validity of the above conclusions about Einstein locality based on equation (\ref{eq:singletzn}) is not limited only to the case of electron singlet system; Instead, they may be generalized and applied to any composite systems. When an $N$-particle composite system ($N\ge2$ is an integer) in an entangled state $|\psi_c\rangle$ is subject to a joint measurement described by operator $\hat{A}_1\hat{A}_2...\hat{A}_N$, any of its member particles $a_k$ ($k=1,2,...,N$) independently undergoes a local $\hat{A}_k$ measurement and thereby is separately projected onto an eigenstate $|\alpha_{lk}\rangle$ ($l=1,2,...$) of the operator $\hat{A}_k$, where $\hat{A}_k|\alpha_{lk}\rangle=\alpha_{lk}|\alpha_{lk}\rangle$ with $\alpha_{lk}$ being a corresponding eigenvalue. In other words, the local measurement 
$\hat{A}_k$ on particle $a_k$ does not change the state of another particle $a_{k'}$ ($k\ne k'$) or affect a local measurement $\hat{A}_{k'}$ on the latter even though the particles are entangled.

The proof of Einstein locality presented here is just the first step to recover the completeness of quantum mechanics. In what follows, as illustrative examples, the developed theory will be applied to the Bell experiments and the double-slit experiments, aiming to provide a unified framework to account for the corresponding results.

\section*{Double-slit experiments with entangled photons}

\begin{figure}[htbp]
\centering
\includegraphics[width=9.2cm]{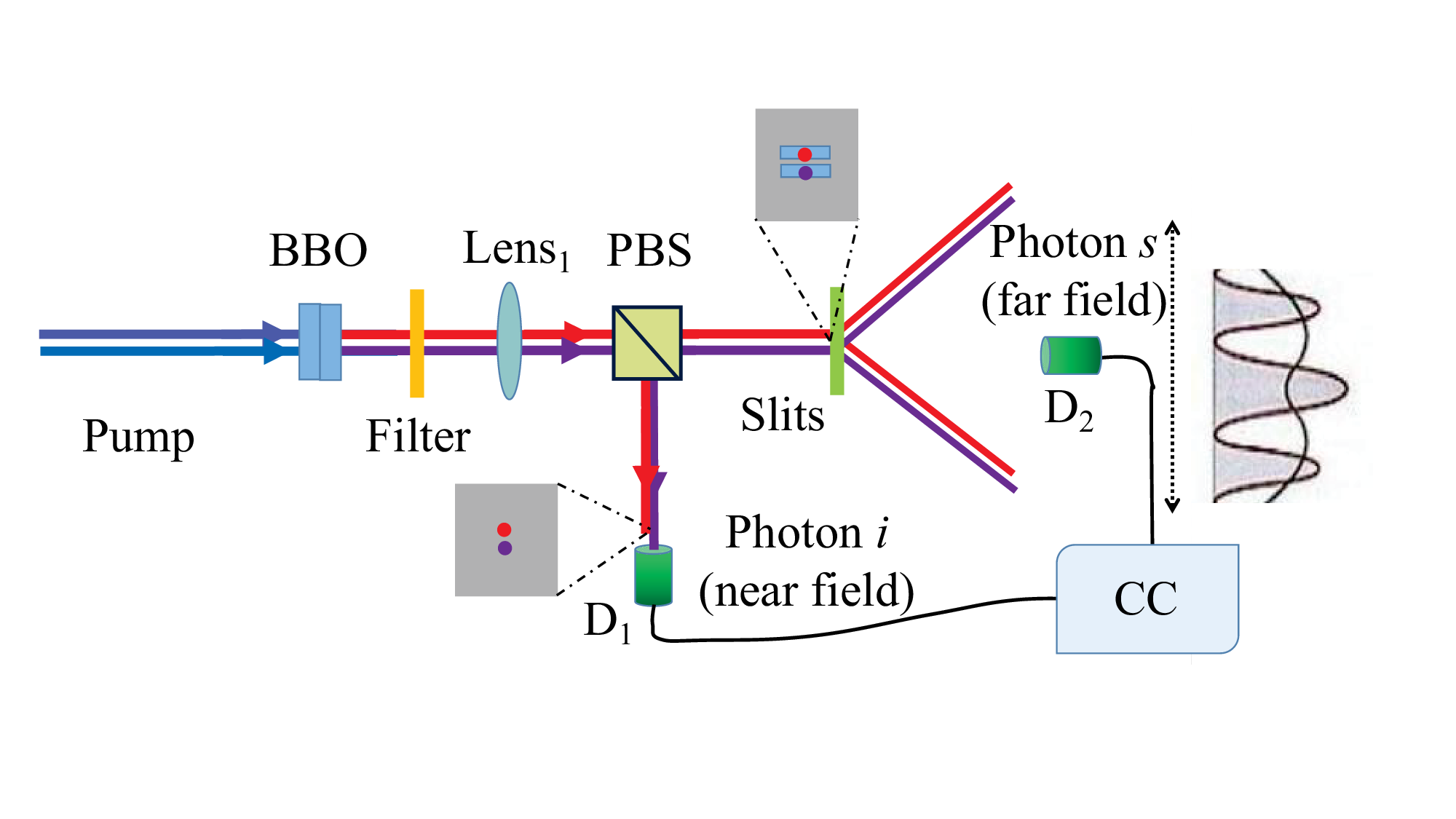}
\caption{A double-slit experiment initially conducted by Menzel {\it et al.} \cite{menzel2012} to investigate wave-particle dualism and the complementarity of entangled photons. The far-field detection of photon $s$ was performed using a D$_2$ detector that was spatially scanned. Curves of different colors represent various intensity maxima in TEM$_{01}$ beam modes. BBO: beta-barium borate crystal. PBS: Polarizing beam splitter. CC: Coincidence counting.}
\label{fig:spdc}
\end{figure}

In view of a relatively light load of mathematical calculations, we will first consider the application of the developed theory to the case of double-slit experiments, for a full understanding of the physical principle that dictates biased sampling in the experimental measurement. Two experiments of such kind were conducted in 2012 and 2013 \cite{menzel2012,menzel2013} to investigate wave-particle dualism and complementarity using entangled photons. In both experiments, the entangled photons ($s$ and $i$) were produced through Type-II spontaneous parametric down-conversion (SPDC) in a beta-barium borate (BBO) crystal (Fig. \ref{fig:spdc}). The unique feature shared by these two experiments was the use of laser pumps in TEM$_{01}$ modes, each displaying two distinct intensity maxima in an upper-lower geometry, thereby generating photon pairs in TEM$_{01}$ modes as a result. Corresponding intensity maxima are represented by the curves of varying colors in Fig. \ref{fig:spdc} (red for the upper maximum and purple for the lower). Only one pair of photons was produced during each measurement time interval ($<$ 5 ns) in each experiment, since the average pump intensity ($\sim$50 mW within a 300 $\mu$m diameter spot) was controlled to be sufficiently low. Paired photons with orthogonal polarizations were then separated by a polarizing beam splitter (PBS). A lens (lens$_1$) was used to couple the generated photons into the opening of a near-field double slit (photon $s$) or directly into a near-field photon detector D$_1$ (photon $i$).

In the reported near-field measurement conditions \cite{menzel2012,menzel2013}, the photons $s$ and $i$ were produced as a composite system in an entangled state described by:
\begin{equation}
|\phi_c\rangle=2^{-1/2}(|u\rangle_s |u\rangle_i + e^{i\theta} |d\rangle_s |d\rangle_i),   \
\label{eq:entphoton}
\end{equation}
where $|u\rangle$ ($|d\rangle$) denotes the state in which a photon is in the upper (lower) intensity maxima of a TEM$_{01}$ mode and the phase $\theta$ is determined by an SPDC process. The quantum correlation represented by equation (\ref{eq:entphoton}) originates from the simultaneous generation of paired photons at the same spatial point. This correlation was experimentally confirmed with 99$\%$ fidelity when both photons were detected in the near fields \cite{menzel2012}. The detector D$_2$ was then moved from the double slit to the far field, to record photon $s$ as a function of its vertical position (see Fig. \ref{fig:spdc}). This was done in coincidence with the photon $i$, registered by detector D$_1$ at the lower intensity maximum of the near-field TEM$_{01}$ mode. The tenants of Einstein non-locality and the observed photon correlation suggested that the detection of photon $i$ at the lower intensity maximum should cause a state collapse in photon $s$ \cite{menzel2012,menzel2013} (i.e., photon $s$ should be present only in the lower intensity maximum, according to equation (\ref{eq:entphoton})). Consequently, photon $s$ was expected to have lost its coherence between the two intensity maxima, thereby preventing the observation of interference fringes when scanning detector D$_2$ in the far field. However, corresponding experimental results supported the opposite explanation. The authors observed far-field interference fringes with a visibility of up to 50\%, which was much higher than the expected 1\% visibility, given the 99\% slit information provided by detector D$_1$ \cite{menzel2012}.  

The presented experimental results \cite{menzel2012,menzel2013}, by invoking the theory of Einstein locality, may then be understood as follows. The local detection of photon $i$ in a $|d\rangle_i$ state did not cause a state collapse for photon $s$ at the near-field double slit, since photon $s$ remained in its initial state. As such, when passing through the double slit, photon $s$ maintained its initial coherence (and necessary power balance) between intensity maxima, in accordance with Einstein locality. It was this coherence, exhibited by photon $s$, that enabled the observation of interference fringes in the far field, signifying Einstein locality as exhibited by the entangled photons (see Methods Mark$_3$ for more detailed analysis).

A deeper understanding of Einstein locality may be achieved in an attempt \cite{Bolduc2014,Leach2016} to explain these experiments using the concept of biased sampling \cite{Giustina2013}. It was proposed \cite{Bolduc2014,Leach2016} that the results reported by Menzel {\it et al.} were a consequence of different sampling techniques respectively used in the measurements of slit and interference fringe information. In other words, the detection of photon $i$ provided slit information in the near field, while the heralded photon $s$ established an interference pattern in the far field. However, a convincing clarification is wanted for the physical principle behind biased sampling in the measurement, given that the concept of biased sampling is inconsistent with that of Einstein non-locality in the interpretation. After all, if Einstein non-locality holds true, the state reduction of photon $i$ should collapse photon $s$ into a $|d\rangle_s$ state according to equation (\ref{eq:entphoton}). Then photon $s$ would lose its coherence between the intensity maxima (the two slits) and no interference could be observed in experiment, which invalidates the proposal of biased sampling. This inconsistency, overlooked so far though, reflects a lack of full understanding of the experimental results at the fundamental level and prevented one from finding the reason why the sampling was biased in the measurement. 

In contrast, the theory of Einstein locality naturally supports the concept of biased sampling for the double-slit experiments. From this theory, the underlying physical principle is as follows: The detection events of the two photons were mutually independent, i.e., the detection of photon $i$ in a $|d\rangle_i$ state did not affect the state of photon $s$ at all despite that the two photons were in an entangled state. Therefore, the latter passing through both slits could stay in its initial state in the near field and hence did not carry much slit information when the visibility of interference fringes reached the maximum, conditioned on the near-field detection of photon $i$ in a $|d\rangle_i$ state, as confirmed by numerical simulations \cite{Bolduc2014}. 

\section*{Bell experiments}
Bell experiments \cite{Freedman1972,Aspect1982,salart2008,Giustina2015,Shalm2015,Hensen2015,bigbell2018} together with Bell's theorem \cite{Bell1964,Clauser1969} have played a very important role in fundamental tests of quantum mechanics; Non-locally correlated behaviors of entangled objects have been firmly verified in experiment. However, to expose the dynamics of Bell non-locality or ``spooky action at a distance" seems a daunting task without introducing Einstein locality. Now we apply the theory of Einstein locality to explain Bell non-locality aiming to reveal the underlying dynamic mechanisms. To this end, we note that Einstein locality has been recognized from equation (\ref{eq:singletzn}), from which Bell non-locality can be derived as well. This equation shows that electrons $a$ and $b$ may be independently projected onto the $|p_{z}\rangle_a$ and $|q_{n}\rangle_b$ states by the $\hat{\sigma}_z^a$ and $\hat{\sigma}_n^b$ measurements, with a joint probability of $2^{-1}\cos^2(\alpha/2)$ or $2^{-1}\sin^2(\alpha/2)$, respectively. Alternatively, the result given by equation (\ref{eq:singletzn}) may also be understood as follows. This equation allows for the prediction of outcomes from $\hat{\sigma}_z^a$ measurements performed on the electron $a$ when it is reduced into a $|p_{z}\rangle_a$ state. Each  $\hat{\sigma}_z^a$ measurement outcome then occurs with a probability of $2^{-1}\cos^2(\alpha/2)$ or $2^{-1}\sin^2(\alpha/2)$, depending on the specific $\hat{\sigma}_n^b$ measurement results from the projection of electron $b$ onto a $|q_{n}\rangle_b$ state. It is this conditional predictability of the measurement results that gives rise to Bell non-locality (see Methods Mark$_4$). 

The fact that both Einstein locality and Bell non-locality can be derived from equation (\ref{eq:singletzn}) reflects an intrinsic consistency between the two notations. This consistency opens a possibility of constructing a theoretical model based on Einstein locality to describes the non-local behaviors of quantum objects observed in Bell experiments and thereby to reveal the dynamics of non-locality, as proved here. To construct such a model, consider an experiment involving an electron singlet system that can be represented as $N_{\pm_{z,n}}^{a,b}$, thereby denoting the number of counted $|\pm_{n,z}\rangle_{a,b}$ electrons in a corresponding $\hat{\sigma}_{z,n}^{a,b}$ measurement. An Einstein locality model described below consists of two core components:\\ 
(1) {\it Coherence kernel}. Each electron is initially in a reduced state governed by a density operator $\hat{\rho}=2^{-1}(|+_n\rangle \langle+_n| + |-_n\rangle\langle-_n|)$ for an arbitrary $\mathbf{n}$-axis. \\
(2) {\it Correlation kernel}. The results of local $\hat{\sigma}_n$ measurements on these electrons are anti-correlated (i.e., $N_{+_n}^a=N_{-_n}^b$ and $N_{-_n}^a=N_{+_n}^b$ for an arbitrary $\mathbf{n}$-axis).

It is worth noting that both the coherence and correlation requirements for the two kernels are naturally satisfied by the electron singlet system and that a single model kernel does not provide the compound system with non-local properties. For instance, the correlation kernel alone guarantees neither Bell non-locality nor Einstein non-locality, as it is trivial to implement $N_{\pm_n}^a=N_{\mp_n}^b$ for arbitrary $\mathbf{n}$-axes in the context of classical optics \cite{Freedman1972,cramer1986}. Interestingly, the combination of two kernels is nonetheless sufficient for explaining a Bell experiment. The key point is to identify some coherence property exhibited by electron $a$ (or electron $b$), which may be achieved by considering two measurements on the particle (i.e., a $\hat{\sigma}_z^a$ measurement and a $\hat{\sigma}_n^a$ measurement). It is not difficult to then calculate the expectation value of the $\hat{\sigma}_z^a\hat{\sigma}_n^a$ operator from the coherence kernel:
\begin{eqnarray}
<\hat{\sigma}_z^a\hat{\sigma}_n^a>
&=& \cos\alpha\ ,
\label{eq:spinave}
\end{eqnarray}
which demonstrates some coherence between the $|p_z\rangle_a$ and $|q_n\rangle_a$ states for electron $a$ (for detailed calculations, see Methods Mark$_5$; the same is true for electron $b$). The $<\hat{\sigma}_z^a\hat{\sigma}_n^a>$ term in equation (\ref{eq:spinave}) could be measured experimentally if $\hat{\sigma}_z^a$ and $\hat{\sigma}_n^a$ measurements could be conducted simultaneously on electron $a$ in its initial state. These measurements would produce experimental data for $N_{p_z}^a$ and $N_{q_n}^a$, which could then be used to calculate a normalized coincidence counting rate, $C(N_{p_z}^a,N_{q_n}^a)$, as a function of $N_{p_z}^a$ and $N_{q_n}^a$. This would then produce an experimentally measured value for $<\hat{\sigma}_z^a\hat{\sigma}_n^a>$, given by \cite{Clauser1969}:
\begin{eqnarray}
<\hat{\sigma}_z^a\hat{\sigma}_n^a>
&=&  \sum_{p,q} \mbox{sign}(pq) \ C(N_{p_z}^a,N_{q_n}^a) \ ,
\label{eq:jointaaa}
\end{eqnarray}
where $\mbox{sign}(\pm\pm)=1$ and $\mbox{sign}(\pm\mp)=-1$. 

\begin{figure}[htbp]
\centering
\includegraphics[width=8cm]{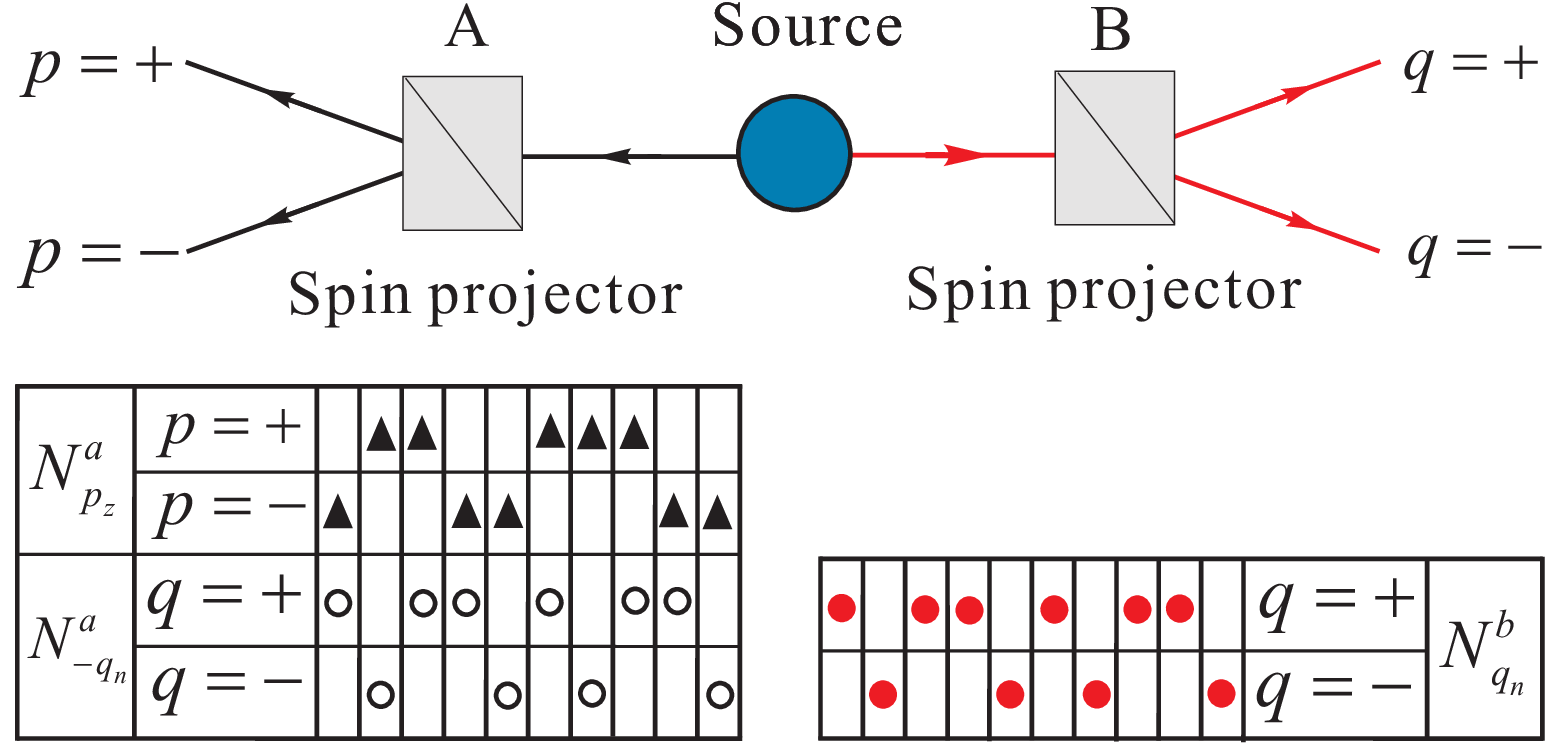}
\caption{A Bell experiment involving an electron singlet system from the perspective of the Einstein locality model. The $N_{p_z}^a$ data (triangles) could be acquired but electron $a$ would be destroyed after being detected in a $|p_z\rangle_a$ state by the $\hat{\sigma}_z^a$ measurement. The $N_{-q_n}^a$ data (open circles) could not be acquired by a $\hat{\sigma}_n^a$ measurement on electron $a$, since it cannot be counted twice in its initial state. The key to acquiring $N_{-q_n}^a$ data is to use the $N_{q_n}^b$ data (red dots) as replicas of the required $N_{-q_n}^a$ samples, since $N_{-q_n}^a=N_{q_n}^b$.}
\label{fig:cc}
\end{figure}

However, it is not possible in practice to simultaneously perform the $\hat{\sigma}_z^a$ and $\hat{\sigma}_n^a$ measurements on electron $a$ in the same initial state. Therefore, the experimental results of relevance to equation (\ref{eq:jointaaa}) are unavailable under typical experimental conditions. Fortunately, photon correlation makes it possible to experimentally achieve the result shown in equation (\ref{eq:jointaaa}) for singlet electron systems. For example, the $N_{p_z}^a$ data may be acquired directly by performing $\hat{\sigma}_z^a$ measurements on electron $a$ in its initial state. Similarly, the $N_{q_{n}}^{a}$ data may be obtained by first acquiring $N_{q_{n}}^{b}$ through an independent $\hat{\sigma}_n^b$ measurement on electron $b$ in its initial state (Einstein locality). Since $N_{-q_n}^a=N_{q_n}^b$ (the correlation kernel), the $N_{q_{n}}^{b}$ data can be used as replicas of $N_{-q_{n}}^{a}$ data that are otherwise unavailable but required by equation (\ref{eq:jointaaa}) (see Fig. \ref{fig:cc}). Due to spin anti-correlation between the electrons, the result achieved by equation (\ref{eq:jointaaa}), using the $N_{p_{z}}^{a}$ and $N_{-q_{n}}^{a}$ data, must be multiplied by a factor of $-1$ when compared with the expectation value from equation (\ref{eq:spinave}). Otherwise, this predicts an experimental value of $<\hat{\sigma}_z^a\hat{\sigma}_n^b>$, which is exactly the negation of $<\hat{\sigma}_z^a\hat{\sigma}_n^a>$ (i.e., $<\hat{\sigma}_z^a\hat{\sigma}_n^b>=-<\hat{\sigma}_z^a\hat{\sigma}_n^a>=-\cos\alpha$, according to equation (\ref{eq:spinave})). It is then trivial to demonstrate a violation of the Bell inequality following a process given by Bell \cite{Bell1964} (see Methods Mark$_6$ for an in-depth theoretical proof). 

\begin{figure*}[htbp]
\centering
\includegraphics[width=16cm]{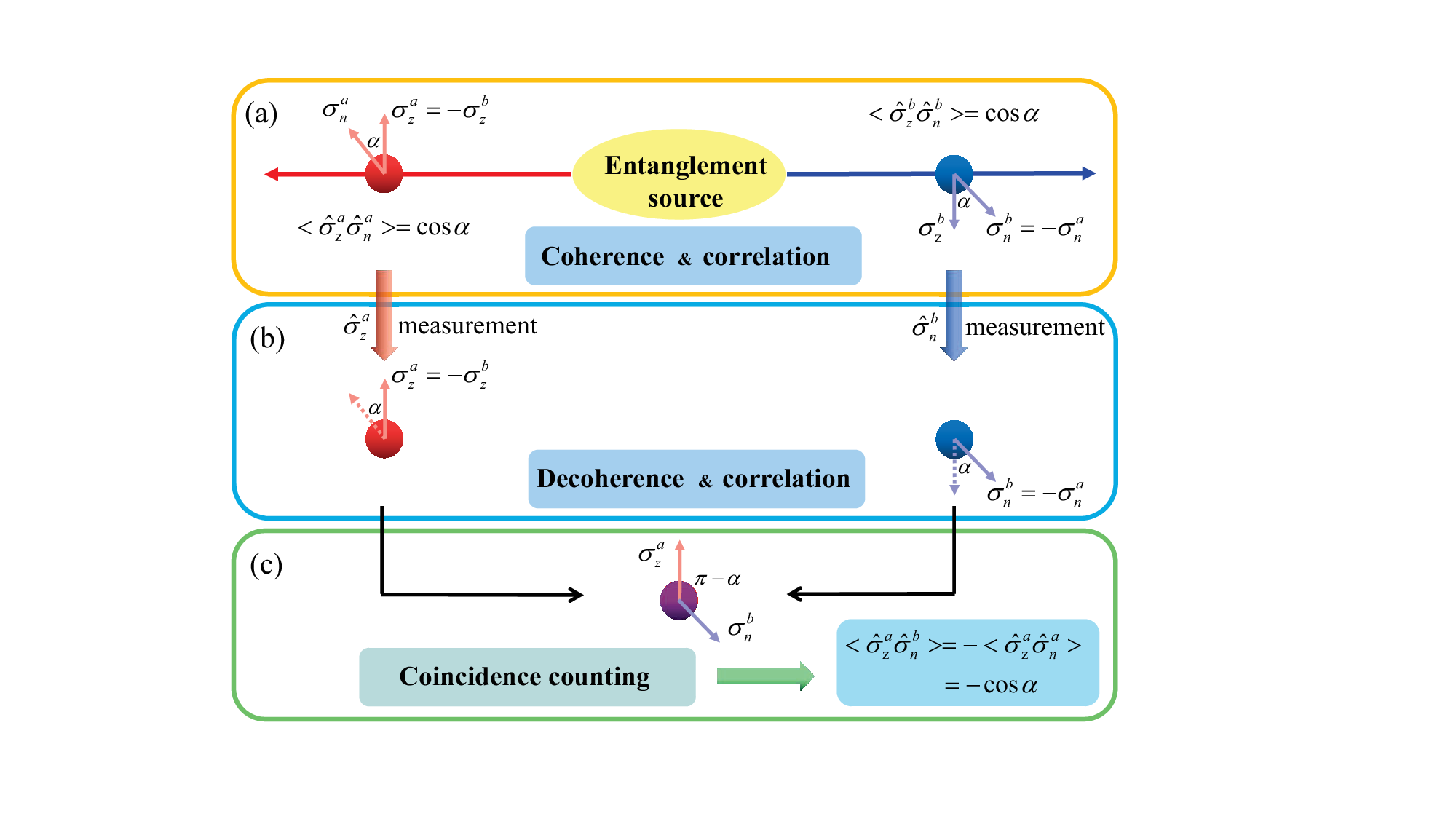}
\caption{An illustration of the dynamic mechanisms for non-locality, as observed in a Bell experiment using an electron singlet system. (\textbf{a}) Entangled electrons produced with a coherence inherent to each particle (given by $<\hat{\sigma}^{a}_z\hat{\sigma}^a_n>=<\hat{\sigma}^{b}_z\hat{\sigma}^b_n>=\cos\alpha$) and a distinct correlation between each particle (given by $\sigma^a_z=-\sigma^b_z$ and $\sigma^b_n=-\sigma^a_n$). (\textbf{b}) After conducting $\hat{\sigma}^{a}_z$ and $\hat{\sigma}^a_n$ measurements, both electrons ($a$ and $b$) experience decoherence and experimental data retaining the correlation information can be acquired for $\sigma^a_z=-\sigma^b_z$ and $\sigma^b_n=-\sigma^a_n$, respectively. (\textbf{c}) Coincidence counting, which automatically exploits the correlation $\sigma^b_n=-\sigma^a_n$ and produces a result of $<\hat{\sigma}^{a}_z\hat{\sigma}^b_n>=-\cos\alpha$, which is the negation of $<\hat{\sigma}^{a}_z\hat{\sigma}^a_n>=\cos\alpha$, due to spin anti-correlation. 
}
\label{fig:nlpc}
\end{figure*}

Bell non-locality is thereby well explained by the Einstein locality model, which reveals the dynamics of Bell non-locality as illustrated in Fig. \ref{fig:nlpc}. In the model, two critical elements of the mechanisms governing electron singlet systems include the coherence exhibited by electrons and their spin anti-correlation. One of these elements alone does not ensure non-locality, violating Bell's inequality, but interestingly the combination of the two does so. From the viewpoint of classical science \cite{Bell1964}, entangled objects exhibit non-local correlations that violate Bell's inequality, as if there were some real ``physical influence faster than the speed of light". On the other hand, from the perspective of quantum science, the behavior of these objects can be fully explained by the Einstein locality model, in which there is little room for ``spooky action at a distance".

\section*{Discussions}
The developed theory, which certifies that Einstein locality is an element of quantum mechanics, has revealed the underlying principle of Bell non-locality in the Bell experiments and that of biased sampling in measurement in the double-slit experiments. Generally speaking, the theory is applicable to any experiments involving quantum entanglement and thereby more unknown physics could be brought into light. In addition, the theory of Einstein locality may be explored to establishes a more solid foundation for the basic concepts of non-locality, entanglement \cite{Almeida2007,horodecki2009,Tendick2020}, and quantum steering \cite{Schrodinger1935,Walborn2011,reid2019,uola2020}, which may provide a novel description of other quantum measurements \cite{Braginsky1992,Wiseman2009,Busch2016} and decoherence \cite{Zurek2003,Wilkinson2020}. This could also facilitate an encouraging approach toward hardware optimization for quantum computers \cite{sjtu2021} (see Methods Mark$_7$).

In the application of the developed theory, of essence is to be clearly aware of a conceptual distinction between Einstein non-locality and Bell non-locality, which are commonly considered to be different manifestations of the same phenomena \cite{Giustina2013,reid2019,uola2020}. Einstein non-locality has often been utilized to provide a physical description explaining the underlying principle of Bell non-locality \cite{cramer1986,wiseman2007}, which suggests that Bell experiments have verified not only Bell non-locality, but also Einstein non-locality. This can be misleading, since Bell did not explicitly distinguish the two types of non-locality, as he initially intended to derive the Bell inequality for experimental tests of the EPR argument \cite{Bell1964}. 

If not clarified, the confusion between Bell non-locality and Einstein non-locality could lead to invalid concepts or inaccurate models in theoretical analysis of experimental observations. For example, in the double-slit experiments, violation of duality was claimed \cite{menzel2012,menzel2013,Bolduc2014} from the experimental results due to the assumption of Einstein non-locality. However, the concept of Einstein locality, and biased sampling as well, suggests that one photon (photon $s$) was observed as ``wave" while its entangled partner (photon $i$) observed as ``particle"; no photon was really observed as ``wave" and ``particle" simultaneously, which in fact did not violate the principle of complementarity. In addition, this confusion would also produce imperceptible obstacles for solving important problems of modern physics, e.g., determining whether quantum many-body entanglement is more fundamental than quantum fields \cite{sjtu2021}.

In summary, we have discovered Einstein locality, a core element of quantum mechanics, that was widely ignored in practical research activities involving quantum entanglement; Unawareness of this element has ruined the completeness of the quantum theory. The theoretical completeness may be recovered by a theory of Einstein locality developed in accordance to the strict formalism of quantum mechanics, using an electron singlet system as an illustrative example. The developed theory can provide a unified framework to account for the results of Bell experiments and double-slit experiments on wave-particle duality using entangled photons. By introducing Einstein locality, the dynamics of Bell non-locality has been exposed and the physical principle behind biased sampling in measurement in the double-slit experiments has been revealed as well. It has been proved that ignorance of Einstein locality in quantum mechanics has led to misinterpretation of important experimental results, such as claim of duality violation in the double-slit experiments. This study is anticipated to not only impact the field of quantum science, by providing more solid basis for other fundamental concepts in quantum theory (e.g., non-locality, entanglement, quantum steering, decoherence, and quantum measurement), but also advance quantum technology with a potential method for optimization of quantum computing hardware.

\section*{Methods}
\subsection*{Wave function decomposition and Einstein locality}
{\bf Mark$_1$}: Equation (\ref{eq:singletzn}) cannot be expanded as a Schmidt decomposition of $|\phi_c\rangle$ for a joint $\hat{\sigma}_z^a\hat{\sigma}_n^b$ measurement on an electron singlet system. This can be understood by investigating how the superposition principle must be applied in quantum theory when expressing the wave function of a system in a superposed form. According to quantum mechanics, if a single particle in a state denoted by a wave function $|\psi\rangle$ is measured to produce an observable $A$, described by an operator $\hat{A}$, the superposition principle may be invoked to express $|\psi\rangle$ in a superposed form as: 
\begin{equation}
|\psi\rangle=\sum_{j=1} a_j |a_j\rangle,\
\label{eq:superposition_1}
\end{equation}
where the complex numbers $a_j$ satisfy $\sum_j|a_j|^2=1$ and $|a_{j}\rangle$ are the eigenstates of the operator $\hat{A}$. A vector $\mathbf{A}$ ($\mathbf{A}^T\equiv \left[a_1, a_2, ..., a_j, ...\right]$) may then be constructed from equation (\ref{eq:superposition_1}) as a representation of the $|\psi\rangle$ state in a Hilbert space spanned by the ${|a_j\rangle}$ bases. In this way, the vector $\mathbf{A}$ fully dictates the projection of the particle onto the final $|a_j\rangle$ state using the $\hat{A}$ measurement. Otherwise, if the particle is measured for an observable $B$ described by an operator $\hat{B}\ne\hat{A}$, this can be expressed as:
\begin{equation}
|\psi\rangle=\sum_{j=1} b_j |b_j\rangle,\
\label{eq:superposition_2}
\end{equation}
in which $\sum_j|b_j|^2=1$ and $|b_{j}\rangle$ are eigenstates of the operator $\hat{B}$. In this case, a constructed vector $\mathbf{B}$ ($\mathbf{B}^T\equiv \left[b_1, b_2, ..., b_j, ...\right]$) governs the measurement-induced projections of the particle onto the final $|b_j\rangle$ states.

Equations (\ref{eq:superposition_1}) and (\ref{eq:superposition_2}) are mathematically equivalent for a given $|\psi\rangle$. However, $\mathbf{A}\ne\mathbf{B}$ generally, as they are distinct representations of the $|\psi\rangle$ state in different Hilbert spaces. Consequently, the ${|a_j\rangle}$ expansion bases must be used for an $\hat{A}$ measurement on the particle when representing $|\psi\rangle$ in a superposed form, such that the outcomes may be appropriately predicted by the vector $\mathbf{A}$. Otherwise, if the ${|b_j\rangle}$ bases are selected when applying the superposition principle to the $|\psi\rangle$ state, the theoretical prediction given by vector $\mathbf{B}$ will not agree with the actual $\hat{A}$ measurement results. Therefore, the choice of expansion bases (CEB) for $|\psi\rangle$ must be measurement oriented whenever the superposition principle is invoked. This is referred to as the CEB rule and must be strictly followed in quantum theory. 
 
Generalizing the CEB rule to composite systems naturally leads to Einstein locality in quantum entanglement. To illustrate, consider a compound system consisting of two particles, $a$ and $b$. Suppose an $\hat{A}$ measurement is performed on particle $a$ and a $\hat{B}$ measurement on particle $b$. The wave function $|\psi_c\rangle$ for this system can then be written as a superposition of the $|a_i\rangle_a |b_j\rangle_b$ states as follows:
\begin{equation}
|\psi_c\rangle=\sum_{ij=1} \gamma_{ij} |a_i\rangle_a |b_j\rangle_b.\
\label{eq:superpositiontwo}
\end{equation}

A matrix $\mathbf{\Gamma}$ may then be constructed as a representation of the $|\psi_c\rangle$ state in a Hilbert space spanned by the ${|a_i\rangle}_a |b_j\rangle_b$ bases, with $\gamma_{ij} = (_a\langle a_i| _b\langle b_j|)|\psi_c\rangle$ denoting the matrix elements used to fully describe measurement-induced projections of the compound system onto the final $|a_i\rangle_a |b_j\rangle_b$ states, with probabilities of $|\gamma_{ij}|^2$ ($\sum_{ij} |\gamma_{ij}|^2=1$). If a different joint measurement is performed on the system, the wave function $|\psi_c\rangle$ must be expressed as a superposition of expansion bases other than $|a_i\rangle_a |b_j\rangle_b$. This produces a different matrix $\mathbf{\Gamma}'$, used to correctly predict measurement outcomes. The measurement-oriented CEB rule for the superposition principle should also be enforced in the case of composite systems.

Unfortunately, applying the CEB rule to composite systems attracted little attention from researchers (e.g., Einstein and his colleagues) \cite{epr}. This lack of CEB rule utilization may be illustrated by an electron singlet state:
\begin{equation}
|\phi_c\rangle=2^{-1/2}(|+_z\rangle_a |-_z\rangle_b - |-_z\rangle_a |+_z\rangle_b)   \ ,
\label{eq:singletz}
\end{equation}
which is a Schmidt decomposition of $|\phi_c\rangle$. Since $|\pm_z\rangle$ are the eigenstates of the $\hat{\sigma}_z$ operator, equation (\ref{eq:singletz}) naturally meets the requirement of the CEB rule when expressing $|\phi_c\rangle$ in a superposed form with $|\pm_z\rangle_a |\pm_z\rangle_b$ bases for joint $\hat{\sigma}_z^a\hat{\sigma}_z^b$ measurements. Therefore, equation (\ref{eq:singletz}) provides accurate predictions for the corresponding joint measurement outcomes. However, if a $\hat{\sigma}_z^a$ measurement is performed on electron $a$ and a $\hat{\sigma}_n^b$ measurement on electron $b$, the CEB rule dictates that $|\phi_c\rangle$ can no longer be expressed in the form of equation (\ref{eq:singletz}). Instead, it must be expanded as a superposition of $|\pm_z\rangle_a |\pm_n\rangle_b$ states as follows: 
\begin{eqnarray}
|\phi_c\rangle&=&\gamma_{++}\ |+_{z}\rangle_a  |+_{n}\rangle_b + \gamma_{+-}\ |+_{z}\rangle_a  |-_{n}\rangle_b\nonumber \\ 
&& + \gamma_{-+}\ |-_{z}\rangle_a  |+_{n}\rangle_b +\gamma_{--}\ |-_{z}\rangle_a  |-_{n}\rangle_b \ ,
\label{eq:singletznmt}
\end{eqnarray}
which are equivalent to equation (\ref{eq:singletzn}). Here $|\pm_{n}\rangle$ represents the eigenstates of the $\hat{\sigma}_n$ operator, 
$|+_{n}\rangle=\cos(\alpha/2)|+_{z}\rangle+\sin(\alpha/2)e^{i\beta}|-_{z}\rangle$, and $|-_{n}\rangle=-\sin(\alpha/2)e^{-i\beta}|+_{z}\rangle+\cos(\alpha/2)|-_{z}\rangle$. The connection between equations (\ref{eq:singletz}) and (\ref{eq:singletznmt}) is complicated in quantum theory. These expressions are mathematically equivalent yet, at the same time, distinct in physics (due to the CEB rule), as explained below. In addition, equation (\ref{eq:singletznmt}) reduces to equation (\ref{eq:singletz}) when $\alpha = 0$. As such, the former can be considered a generalization of the latter for the generic condition of $\alpha\ne 0$. In this case, two $\mathbf{\Gamma}$ matrices can be constructed from equations (\ref{eq:singletz}) and (\ref{eq:singletznmt}), representing the $|\phi_c\rangle$ state in two Hilbert spaces. Corresponding joint measurement results are respectively given by these two evidently different matrices, from which it follows that equations (\ref{eq:singletz}) and (\ref{eq:singletznmt}) exhibit distinct physical meanings. This explains why equations (\ref{eq:singletzn}) and (\ref{eq:singletznmt}) cannot be expressed as a Schmidt decomposition of $|\phi_c\rangle$ (as with equation (\ref{eq:singletz})) for joint $\hat{\sigma}_z^a\hat{\sigma}_n^b$ measurements of the singlet system when $\alpha\ne 0$. From equation (\ref{eq:singletzn}) or (\ref{eq:singletznmt}), Einstein locality in quantum entanglement can be proved without much difficulty, as done in the main text.

{\bf Mark$_2$}: In practice, representations of equation (\ref{eq:singletz}) as a Schmidt decomposition of $|\phi_c\rangle$ have almost exclusively been used to construct a physical interpretation of the underlying principles of Bell experiments, involving joint $\hat{\sigma}_z\hat{\sigma}_n$ measurements on entangled electrons ($\alpha\ne 0$), which obviously violate the CEB rule. However, equation (\ref{eq:singletz}) alone cannot provide a valid prediction of $\hat{\sigma}_z^a\hat{\sigma}_n^b$ measurement results, which require additional elements to be introduced into the theory to compensate for violating the CEB rule. One of these elements is the concept of Einstein non-locality, which can be traced back to the famous EPR paper \cite{epr}. The use of a superposition principle without following the CEB rule was one of the subtlest weaknesses of the EPR paper, as highlighted in that paper by equations (7) and (8), which did not specify any joint measurements on the composite system of interest \cite{epr}. According to Einstein non-locality, the electrons $a$ and $b$, described by equation (\ref{eq:singletz}), must be simultaneously maintained in $|p_{z}\rangle_a$ and $|-p_{z}\rangle_b$ (or $|-q_{n}\rangle_a$ and $|q_{n}\rangle_b$) states, respectively, after performing a $\hat{\sigma}_z^a$ ($\hat{\sigma}_n^b$) measurement on electron $a$ ($b$). However, Einstein non-locality still lacks experimental validation, since Bell experiments can be satisfactorily explained by assuming Einstein locality, as demonstrated in Fig. \ref{fig:cc} of this work. 

One potential issue with Einstein locality is its compatibility with the uncertainty principle in the case of simultaneous $\hat{\sigma}_z^a$ and $\hat{\sigma}_y^b$ measurements on electrons $a$ and $b$. It could be argued \cite{epr} that arbitrarily precise results for both the $\hat{\sigma}_z^a$ and $\hat{\sigma}_y^b$ measurements are expected from Einstein locality and, at the same time, the $\sigma_y$ value of electron $a$ may be precisely inferred from the $\hat{\sigma}_y^b$ measurement on electron $b$, due to quantum correlation in the singlet system. It then follows that precise $\sigma_y$ and $\sigma_z$ values for electron $a$ violate the uncertainty principle. However, this approach overlooks the fact that, whenever electron $a$ is reduced into a $|+_{z}\rangle_a$ or $|-_{z}\rangle_a$ state by a $\hat{\sigma}_z^a$ measurement, electron $b$ will be randomly projected onto a $|\pm_y\rangle_b$ state with a 50\% probability. This can be achieved using a $\hat{\sigma}_y^b$ measurement, as described by equation (\ref{eq:singletznmt}). In other words, a $\hat{\sigma}_y^b$ measurement on electron $b$ will not produce useful $\sigma_y$ information when electron $a$ is detected in either a $|+_{z}\rangle_a$ or $|-_{z}\rangle_a$ state. As such, the uncertainty principle will not be violated and it is therefore compatible with Einstein locality.

Other interesting results, found in the literature \cite{Griffiths2020,Bedard2021}, may be used to resolve the non-locality issue by applying the concept of locality. For instance, the ``consistent histories" approach suggests that non-locality is inconsistent with Hilbert-space quantum mechanics \cite{Griffiths2020}. However, in this treatment, wave function collapse is considered only as a calculation device that does not correspond to any real physical processes, which does not follow the formalism of quantum mechanics. Since the physical state of a quantum object is fully described by its wave function, the corresponding collapse must be accompanied by a change in the physical state of the object, as required by quantum theory. Another attempt at representing non-locality \cite{Bedard2021} involves analyzing quantum information processing in the Heisenberg picture using super dense coding protocols as a calculation tool for quantum information science, thereby asserting the nature of locality. In the context of locally accessible information, this is simply an alternative way to re-express the results of the no-signaling theorem \cite{Ghirardi1980}, which (unfortunately) does not prove locality.

\subsection*{Result analysis for the double-slit experiments}

{\bf Mark$_3$}: The experimental results reported by Menzel {\it et al.} \cite{menzel2012,menzel2013} may be understood by noting that 1) they are repeatable given that they were conducted with different optical systems leading to the same results and 2) they are not contradictory to other experiments on similar topics \cite{kaiser2012,Bienfait2020}, due to their unique photon detection methods \cite{menzel2012}. The detection of photons should typically be implemented in a manner such that their correlation remains as produced. When being detected, these entangled photons should be projected onto the $|u\rangle_s |u\rangle_i$ or $|d\rangle_s |d\rangle_i$ states with equal probabilities, according to equation (\ref{eq:entphoton}). This quantum correlation between the photons has been observed when they are collected in near fields  \cite{menzel2012,menzel2013}. These two experiments are distinguished from others by their employed measurement techniques \cite{kaiser2012,Bienfait2020}, since the photon $s$ was collected in a {\it far field}, while photon $i$ was detected in a {\it near field}. Under this featured detection condition, the measured photon correlation (given by equation (\ref{eq:entphoton})) is not expected to be retained, since it is invalidated by the CEB rule, due to the far-field detection of photon $s$. This occurs because none of the $|u\rangle_s$ or $|d\rangle_s$ bases for expressing $|\phi_c\rangle$ in the superposed form of equation (\ref{eq:entphoton}) are eigenstates of a far-field measurement operator. The Fourier integral of a near-field measurement operator cannot project photon $s$ onto a $|u\rangle_s$ or $|d\rangle_s$ state in the near field.

It has been suggested \cite{menzel2012} that these observed interference fringes were the result of a TEM$_{01}$ mode function, including the “unoccupied” intensity maximum imprinted on photon $s$. This would have occurred in the form of a superposition of two wave vectors, resulting from a special choice of TEM$_{01}$ pump modes \cite{menzel2012,menzel2013}. However, the selected mode function was only one of the necessary, but not sufficient, conditions for the interference fringes to exist in the far field. Other requirements included coherence and power balance between the two peaks of the TEM$_{01}$ mode. These observed interference fringes signified that both intensity maxima in the TEM$_{01}$ mode must have been ``occupied" by photon $s$, with the necessary coherence and power balance. As such, the detection of photon $i$ could not provide slit information for photon $s$, which remained in its initial $\hat{\rho}$ state at the double slit (Einstein locality). Therefore, neither wave-particle dualism nor complementarity was actually violated in the double-slit experiments. The claim of duality violation was indeed a consequence arising from the assumption of Einstein non-locality by all the relevant investigations \cite{menzel2012,menzel2013,Bolduc2014}.

\subsection*{Theoretical calculations relevant to Bell non-locality}

{\bf Mark$_4$}: In a Bell experiment involving an electron singlet system, on which a joint $\hat{\sigma}_z^a\hat{\sigma}_n^b$ measurement is performed, the expectation value of the $\hat{\sigma}_z^a\hat{\sigma}_n^b$ operator may be simply calculated using equation (\ref{eq:singletzn}) as follows:
\begin{eqnarray}
<\hat{\sigma}_z^a\hat{\sigma}_n^b>&=&\langle\psi_c|\hat{\sigma}_z^a\hat{\sigma}_n^b|\psi_c\rangle \nonumber \\
&=& 2^{-1}\sin^2(\alpha/2){}_a\langle+_z|{}_b\langle+_n|\hat{\sigma}_z^a\hat{\sigma}_n^b|+_z\rangle_a|+_n\rangle_b \nonumber \\
&+&  2^{-1}\cos^2(\alpha/2){}_a\langle+_z|{}_b\langle-_n|\hat{\sigma}_z^a\hat{\sigma}_n^b|+_z\rangle_a|-_n\rangle_b \nonumber \\
&+&2^{-1}\cos^2(\alpha/2){}_a\langle-_z|{}_b\langle+_n|\hat{\sigma}_z^a\hat{\sigma}_n^b|-_z\rangle_a|+_n\rangle_b \nonumber \\
&+&2^{-1}\sin^2(\alpha/2){}_a\langle-_z|{}_b\langle-_n|\hat{\sigma}_z^a\hat{\sigma}_n^b|-_z\rangle_a|-_n\rangle_b  \nonumber\\
&=& \sin^2(\alpha/2)-\cos^2(\alpha/2)   \nonumber\\
&=& - \cos\alpha\ .
\label{eq:singletave}
\end{eqnarray}

These results are a direct consequence of the conditional predictability of measurements made with a probability of $2^{-1}\cos^2(\alpha/2)$ or $2^{-1}\sin^2(\alpha/2)$, depending on the joint $\hat{\sigma}_z^a\hat{\sigma}_n^b$ outcomes. To be specific, the probability that both the $\hat{\sigma}_z^a$ and $\hat{\sigma}_n^b$ measurement results are positive (or negative) is given by $2^{-1}\sin^2(\alpha/2)$, otherwise it is $2^{-1}\cos^2(\alpha/2)$. It is then straightforward to demonstrate a violation of the Bell inequality following a process given by Bell in his 1964 paper \cite{Bell1964}, which signifies Bell non-locality exhibited by the electron singlet system.

{\bf Mark$_5$}: The calculations following equation (\ref{eq:singletave}) do not provide visual insights into the nature of Bell non-locality. In this study, an Einstein locality model is constructed to provide a visual description of the underlying principles governing the Bell experiment, to compensate for a lack of illustration by the equations. This model based on the concept of Einstein locality also aims to expose the dynamical mechanisms governing Bell non-locality. To that end, in the following we illustrate the coherence properties exhibited by electron $a$. In the case of two measurements performed on electron $a$ (a $\hat{\sigma}_z^a$ and a $\hat{\sigma}_n^a$ measurement), the coherence kernel suggests the expectation value of the $\hat{\sigma}_z^a\hat{\sigma}_n^a$ operator can be expressed as:
\begin{eqnarray}
<\hat{\sigma}_z^a\hat{\sigma}_n^a>
&=& \mbox{Tr}(\hat{\rho}\hat{\sigma}_z^a\hat{\sigma}_n^a) \nonumber\\
&=&2^{-1}({}_a\langle+_n| \hat{\sigma}_z^a\hat{\sigma}_n^a |+_n\rangle_a + {}_a\langle-_n| \hat{\sigma}_z^a\hat{\sigma}_n^a |-_n\rangle_a) \nonumber\\
&=& \cos\alpha\ ,
\label{eq:spinavemt}
\end{eqnarray}
in which the condition $\hat{\sigma}_n^a |q_n\rangle_a = q |q_n\rangle_a$ 
is exploited in the second step and $|\pm_n\rangle_a$ is then expanded as the superposition of $|\pm_z\rangle_a$, followed by the use of $\hat{\sigma}_z^a |p_z\rangle_a = p |p_z\rangle_a$ in the calculations. Equation (\ref{eq:spinavemt}) represents the process used to derive equation (\ref{eq:spinave}). Furthermore, inserting the identity operator $\sum_{s}|s_z\rangle\langle s_z|$ ($s=\pm$) into equation (\ref{eq:spinavemt}), between $\hat{\sigma}_z^a$ and $\hat{\sigma}_n^a$, gives:
\begin{eqnarray}
<\hat{\sigma}_z^a\hat{\sigma}_n^a>
&=&  2^{-1}\sum_{p,q} \mbox{sign}(pq) \ |{}_a\langle p_z|q_n\rangle_a|^2
\ .
\label{eq:jointmeas}
\end{eqnarray}

The conditions $\hat{\sigma}_n^a |q_n\rangle_a = q |q_n\rangle_a$ and $\hat{\sigma}_z^a |p_z\rangle_a = p |p_z\rangle_a$ were again exploited in these calculations. The results of equations (\ref{eq:spinavemt}) and (\ref{eq:jointmeas}) may be understood as follows. Electron $a$ can be projected from its initial state ($\hat{\rho}$) onto the eigenstates of $\hat{\sigma}_n^a$ or $\hat{\sigma}_z^a$ using corresponding measurements (i.e., $\hat{\sigma}_n^a: \hat{\rho}\rightarrow|q_n\rangle\langle q_n|$ or $\hat{\sigma}_z^a: \hat{\rho}\rightarrow|p_z\rangle\langle p_z|$). Electron $a$ then exhibits the property of coherence between the $|p_z\rangle_a$ and $|q_n\rangle_a$ states, such that the averaged value of $\mbox{sign}(pq)|{}_a\langle p_z|q_n\rangle_a|^2$ (i.e., the average overlap between the $|p_z\rangle_a$ and $|q_n\rangle_a$ states weighted with $\mbox{sign}(pq)$) is proportional to $\cos\alpha$ (the proportionality constant is usually normalized). The same is true for both electrons, which explains the use of the coherence kernel in the model. 

{\bf Mark$_6$}: In what follows, the combination of two kernels is shown to constitute a sufficient condition for violating the Bell inequality (Bell non-locality). Note that electron $a$ in the $\hat{\rho}$ state may be projected either onto a $|p_z\rangle$ state with a $p$ outcome (by a $\hat{\sigma}_z^a$ measurement) or onto a $|q_n\rangle$ state with a $q$ outcome (by a $\hat{\sigma}_n^a$ measurement). Also, the expectation value for the $\hat{\sigma}_z^a\hat{\sigma}_n^a$ operator may be determined from equation (\ref{eq:jointmeas}), where the $|{}_a\langle p_z|q_n\rangle_a|^2$ term (representing the intersection of the $|p_z\rangle_a$ and $|q_n\rangle_a$ states) is proportional to the probability that ``electron $a$ exhibits a theoretically expected $p$ result for the $|p_z\rangle_a$ state in the $\hat{\sigma}_z^a$ measurement and a theoretically expected $q$ result for the $|q_n\rangle_a$ state in the $\hat{\sigma}_n^a$ measurement". This experimental probability can be characterized by the normalized coincidence counting rate $C(N_{p_z}^a,N_{q_n}^a)$ \cite{Clauser1969}. Consequently, $C(N_{p_z}^a,N_{q_n}^a)=2^{-1}|{_a}\langle p_z|q_n\rangle_a|^2$, with $2^{-1}$ serving as a normalization coefficient. Inserting this expression into equation (\ref{eq:jointmeas}) gives equation (\ref{eq:jointaaa}). Invoking the correlation kernel in the model facilitates the substitution of $N_{q_n}^a=N_{-q_n}^b$ into equation (\ref{eq:jointaaa}), which yields: 
\begin{eqnarray}
<\hat{\sigma}_z^a\hat{\sigma}_n^a>
&=&  -\sum_{p,q} \mbox{sign}(pq) \ C(N_{p_z}^a,N_{q_n}^b) \nonumber\\
&=& -<\hat{\sigma}_z^a\hat{\sigma}_n^b>\ ,
\label{eq:jointaba}
\end{eqnarray}
in which $q\rightarrow -q$ was employed. The final step originates from the standard method used to obtain the value of $<\hat{\sigma}_z^a\hat{\sigma}_n^b>$ experimentally \cite{Clauser1969}, in which the normalized coincidence counting rate $C(N_{p_z}^a,N_{q_n}^b)$ is equal to the probability that this measurement produces $p$ and $q$ for the electrons ($a$ and $b$) in the joint $\hat{\sigma}_z^a\hat{\sigma}_n^b$ measurement, respectively. Plugging equation (\ref{eq:spinavemt}) into equation  (\ref{eq:jointaba}) leads to equation (\ref{eq:singletave}), demonstrating the violation of a Bell inequality (Bell non-locality) for a Bell experiment in which a joint $\hat{\sigma}_z^a\hat{\sigma}_n^b$ measurement is performed on an electron singlet system, according to the process provided by Bell \cite{Bell1964}. In the proof of Bell non-locality, equation (\ref{eq:spinavemt}) is derived from the coherence kernel in the model and equation (\ref{eq:jointaba}) is developed from the correlation kernel. Although each kernel is local in the context of Einstein locality, a combination of the two kernels ensures Bell non-locality, as demonstrated by the violation of the Bell inequality.

\subsection*{Impacts on quantum science and technology}

Mark$_7$: The concepts of entanglement and quantum steering were introduced into quantum theory by Schr$\ddot{\mbox{o}}$dinger \cite{Schrodinger1935} as a provoked response to his interest in the EPR paper \cite{epr}. The quantum steering mechanism introduced by Schr$\ddot{\mbox{o}}$dinger involved one particle influencing the wave function of its entangled partner at a distance, through a suitable measurement performed on the former. It follows from Einstein locality that the essence of entanglement and quantum steering lies in the conditional predictability of experimental outcomes that depend on specific joint measurements in compound systems. Obviously, Schr$\ddot{\mbox{o}}$dinger's original understanding \cite{Schrodinger1935} of entanglement and quantum steering conflicted with Einstein locality. Therefore, Schr$\ddot{\mbox{o}}$dinger's version of these concepts must be distinguished from their modern descriptions \cite{wiseman2007}, such that these two important effects may be consolidated. Since Bell non-locality ensures the viability of each \cite{wiseman2007}, the Einstein locality model is sufficient to explain not only Bell non-locality, but also entanglement and quantum steering in their modern forms \cite{wiseman2007,reid2019,uola2020}. 

In addition, the discovery of Einstein locality may provide valuable clues for potentially new science describing quantum measurements \cite{Braginsky1992,Wiseman2009,Busch2016,Pan2019} and decoherence \cite{Zurek2003,Wilkinson2020}, two closely-related subjects of relevance to non-locality. New treatments of quantum measurement problems \cite{Braginsky1992,Wiseman2009,Busch2016} are thus needed to integrate the idea of Einstein locality into a relevant theory. Similarly, the destruction of coherence between system states associated with observables sensed in the environment \cite{Zurek2003,Almeida2007} is of great significance to the foundations of quantum science and engineering. It is believed that a system in a ``Schr$\ddot{\mbox{o}}$dinger's cat" state is extremely fragile because of decoherence. However, the Einstein locality model may provide a new perspective on this topic. To illustrate, consider a multi-particle system of three photons ($a$, $b$, and $c$) in an entangled state \cite{Greenberger1990,Bouwmeester1999} given by:
\begin{equation}
|\phi_{abc}\rangle=2^{-1/2}(|h\rangle_a   |h\rangle_b   |h\rangle_c + |v\rangle_a  |v\rangle_b  |v\rangle_c)   \ ,
\label{eq:singletz3}
\end{equation}
where $|h\rangle$ and $|v\rangle$ represent horizontal and vertical polarization states, respectively. Equation (\ref{eq:singletz3}) indicates that such a system is subject to a joint linear-polarization measurement, as required by the CEB rule. It is then trivial to show that a reduced density operator for each photon can be expressed as $\hat{\rho}=2^{-1}(|h\rangle \langle h| + |v\rangle\langle v|)$. The Einstein locality model for the electron singlet state can then be generalized to the case of three-photon entanglement with a simple modification to the correlation kernel: $N_{h}^a=N_{h}^b=N_{h}^c$ and $N_{v}^a=N_{v}^b=N_{v}^c$. In other words, the results of local polarization measurements for photons along the horizontal and vertical directions are correlated. In this notation, $N_{h,v}^{a,b,c}$ denotes the number of counted $|h\rangle$ or $|v\rangle$ photons in corresponding measurements. 

\begin{figure}[htbp]
\centering
\includegraphics[width=8.5cm]{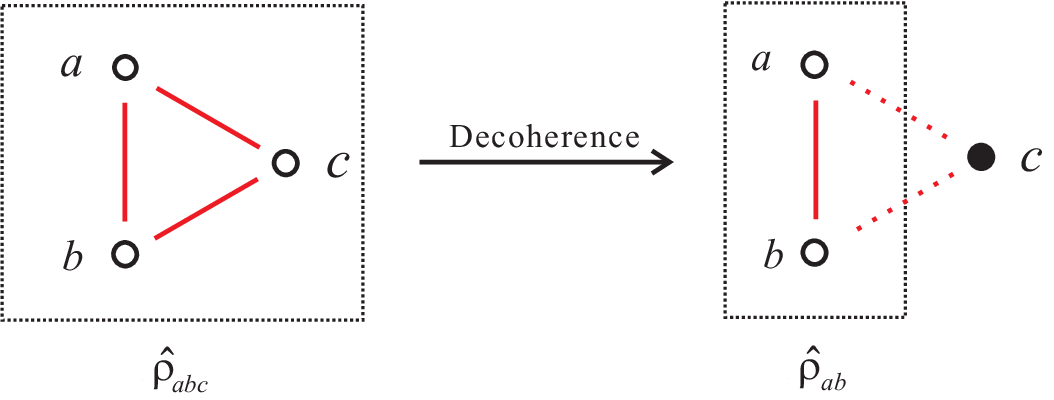}
\caption{A typical decoherence process in the Einstein locality model. A single three-photon system ($a$, $b$, and $c$) undergoes decoherence and hence a state collapse. This system is initially in a $\hat{\rho}_{abc}=|\phi_{abc}\rangle\langle\phi_{abc}|$ state and, after undergoing decoherence, a subsystem of photons $a$ and $b$ is produced in a $\hat{\rho}_{ab}$ state, in which the two photon states and their corresponding correlation remain preserved. Each circle represents an initial $\hat{\rho}$ state for a photon, while the black dot indicate a collapsed state. The red lines describe photon correlations and the red dashed lines denote collapsed correlations.}
\label{fig:decoh}
\end{figure}

Suppose one of these three photons (e.g., photon $c$) suffers from decoherence and hence a state collapse due to the environment. The Einstein locality model then requires the states of photons $a$ and $b$ and the correlation between them to remain unchanged despite the state collapse of photon $c$ (see Fig. \ref{fig:decoh}). It immediately follows that the subsystem of photons $a$ and $b$ must be in a state of $\hat{\rho}_{ab}\ne \hat{\rho}^{\ r}_{ab}$, where $\hat{\rho}^{\ r}_{ab}=2^{-1}(|h\rangle_a |h\rangle_b\langle h|_b\langle h|_a + |v\rangle_a |v\rangle_b\langle v|_b\langle v|_a)$ denotes a mixed state in which the photons are completely uncorrelated. In accordance with Einstein locality and the intact correlation between photons 
$a$ and $b$, it may be assumed that these two photons are likely in an entangled state given by $\hat{\rho}_{ab}=|\phi_{ab}\rangle\langle \phi_{ab}|$, where $|\phi_{ab}\rangle$ can be expressed as:
\begin{equation}
|\phi_{ab}\rangle=2^{-1/2}(|h\rangle_a |h\rangle_b + e^{i\phi} |v\rangle_a |v\rangle_b)   \ ,
\label{eq:singletz2}
\end{equation}
in which $\phi$ is an arbitrary phase to be determined experimentally.

The experimental validation of equation (\ref{eq:singletz2}) may help to provide a novel approach to hardware optimization in quantum computing. To illustrate, suppose a computing task is assigned to a two-qubit quantum computer constructed from photons $a$ and $c$, with photon $b$ acting as a backup qubit. In the case that photon $c$ suffers from decoherence due to the environment, the computer may remain on task after the backup qubit (photon $b$) is implemented as a substitute for photon $c$. If this strategy is successfully adopted in practice, the stringent requirements for protecting all qubits in a quantum computer from decoherence could be substantially relaxed, which would benefit ongoing research efforts to construct commercial quantum computers. While the validity of equation (\ref{eq:singletz2}) is still subject to experimental justification, potentially new science describing decoherence cannot be denied by preconceptions. In any experiment designed to test the validity of equation (\ref{eq:singletz2}), the value of the unknown phase angle $\phi$ must be determined and then compensated for at the beginning of the experiment. In addition, one must note that Einstein locality forbids any feedback information from the collapsed photon $c$, during a measurement of the quantum states for photons $a$ and $b$. Specifically, the detection of photon $c$ cannot be used as a triggering signal for the acquisition of $N_{h,v}^{a,b}$ data.

\section*{Acknowledgments}
This work was financially supported by the National Natural Science Foundation of China (Grant No. 12074110). The author is grateful to K. Wu and L. Chen for preparing some of the figures.

\section*{Competing interests}
The author declares he has no competing interests.

\section*{Author contributions}
S.F. contributed to all aspects of the manuscript.

\section*{Data and materials availability}
All data are available in the main text or the supplementary materials.





\clearpage

\section*{Figure legends}
Figure 1. A double-slit experiment initially conducted by Menzel {\it et al.} \cite{menzel2012} to investigate wave-particle dualism and the complementarity of entangled photons. The far-field detection of photon $s$ was performed using a D$_2$ detector that was spatially scanned. Curves of different colors represent various intensity maxima in TEM$_{01}$ beam modes. BBO: beta-barium borate crystal. PBS: Polarizing beam splitter. CC: Coincidence counting.

Figure 2. A Bell experiment involving an electron singlet system from the perspective of the Einstein locality model. The $N_{p_z}^a$ data (triangles) could be acquired but electron $a$ would be destroyed after being detected in a $|p_z\rangle_a$ state by the $\hat{\sigma}_z^a$ measurement. The $N_{-q_n}^a$ data (open circles) could not be acquired by a $\hat{\sigma}_n^a$ measurement on electron $a$, since it cannot be counted twice in its initial state. The key to acquiring $N_{-q_n}^a$ data is to use the $N_{q_n}^b$ data (red dots) as replicas of the required $N_{-q_n}^a$ samples, since $N_{-q_n}^a=N_{q_n}^b$.

Figure 3. An illustration of the dynamic mechanisms for non-locality, as observed in a Bell experiment using an electron singlet system. (a) Entangled electrons produced with a coherence inherent to each particle (given by $<\hat{\sigma}^{a}_z\hat{\sigma}^a_n>=<\hat{\sigma}^{b}_z\hat{\sigma}^b_n>=\cos\alpha$) and a distinct correlation between each particle (given by $\sigma^a_z=-\sigma^b_z$ and $\sigma^b_n=-\sigma^a_n$). (b) After conducting $\hat{\sigma}^{a}_z$ and $\hat{\sigma}^a_n$ measurements, both electrons ($a$ and $b$) experience decoherence and experimental data retaining the correlation information can be acquired for $\sigma^a_z=-\sigma^b_z$ and $\sigma^b_n=-\sigma^a_n$, respectively. (c) Coincidence counting, which automatically exploits the correlation $\sigma^b_n=-\sigma^a_n$ and produces a result of $<\hat{\sigma}^{a}_z\hat{\sigma}^b_n>=-\cos\alpha$, which is the negation of $<\hat{\sigma}^{a}_z\hat{\sigma}^a_n>=\cos\alpha$, due to spin anti-correlation.

Figure 4. A typical decoherence process in the Einstein locality model. A single three-photon system ($a$, $b$, and $c$) undergoes decoherence and hence a state collapse. This system is initially in a $\hat{\rho}_{abc}=|\phi_{abc}\rangle\langle\phi_{abc}|$ state and, after undergoing decoherence, a subsystem of photons $a$ and $b$ is produced in a $\hat{\rho}_{ab}$ state, in which the two photon states and their corresponding correlation remain preserved. Each circle represents an initial $\hat{\rho}$ state for a photon, while the black dot indicate a collapsed state. The red lines describe photon correlations and the red dashed lines denote collapsed correlations.



\end{document}